\documentclass[12pt]{article}
\usepackage{amsmath}
\def\be{\begin{equation}}
\def\ee{\end{equation}}
\def\arr{\begin{array}{rll}}
\def\ea{\end{array}}
\def\bea{\begin{eqnarray}}
\def\eea{\end{eqnarray}}

\def\N2{$N{=}2$}

\def\>{\rangle}
\def\<{\langle}
\def\+{\dagger}
\def\={\ =\ }

\def\bal{\begin{aligned}}
\def\eal{\end{aligned}}
\begin{document}
\begin{titlepage}
\setcounter{page}{0}
\begin{center}
{\LARGE\bf Some metrics admitting nonpolynomial first }\\
\vskip 0.5cm
{\LARGE\bf integrals of the geodesic equation }\\
\vskip 1.5cm
\textrm{\Large Anton Galajinsky \ }
\vskip 0.7cm
{\it
Tomsk Polytechnic University, 634050 Tomsk, Lenin Ave. 30, Russia} \\
\vskip 0.2cm

\vskip 0.2cm
{e-mail: galajin@tpu.ru}
\vskip 0.5cm
\end{center}

\begin{abstract} \noindent
It is commonly known that Killing vectors and tensors are in one--to--one correspondence with polynomial first integrals of the geodesic equation.
In this work, metrics admitting nonpolynomial first integrals of the geodesic equation are constructed, each of which revealing a chain of generalised Killing vectors.
\end{abstract}

\vspace{0.5cm}

\end{titlepage}
\renewcommand{\thefootnote}{\arabic{footnote}}
\setcounter{footnote}0

\noindent
{\bf 1. Introduction}\\

\noindent
It is commonly known that symmetries of a spacetime generated by Killing vectors or tensors are in one--to--one correspondence with polynomial first integrals of the geodesic equation. This is most easily seen by making recourse to the Hamiltonian formalism.
Introducing canonical pairs $(x^i,p_i)$, with $i=1,\dots,d$, which obey the Poisson bracket $\{x^i,p_j \}={\delta_j}^i$,  the geodesic Hamiltonian\footnote{Throughout the paper, summation over repeated indices is understood unless otherwise is stated explicitly.} $H=\frac 12 g^{ij}(x) p_i p_j$, where $g^{ij}(x)$ is the inverse of a covariantly constant metric tensor $g_{ij}(x)$, and a monomial $\xi^{i_1 \dots i_n}(x) p_{i_1} \dots p_{i_n}$ involving a symmetric tensor field $\xi^{i_1 \dots i_n}(x)$, one readily gets
\be\label{conn}
\{\xi^{i_1 \dots i_n}(x) p_{i_1} \dots p_{i_n},H \}=\nabla^{i_1} \xi^{i_2 \dots i_{n+1}}(x) p_{i_1} \dots p_{i_{n+1}},
\ee
where $\nabla^i$ is the covariant derivative. If $\xi^{i_1 \dots i_n}(x)$ obeys Killing's equation, $\nabla^{(i_1} \xi^{i_2 \dots i_{n+1})}(x)=0$, then $\xi^{i_1 \dots i_n}(x) p_{i_1} \dots p_{i_n}$ is a constant of the motion of the geodesic equation, and vice versa.

The importance of the interrelation above is hard to overestimate.
It gives a clue for establishing the complete integrability of the geodesic equation formulated in various black hole spacetimes, as well as allows one to separate variables in the Hamilton--Jacobi, Klein--Gordon and Dirac equations in strong gravitational fields.\footnote{ There is a vast literature on the subject. For a comprehensive  recent account and further references see \cite{FKK}.} Worth mentioning also is a considerable body of recent work on general relativistic description of integrable systems with finitely many degrees of freedom \cite{GHKW}--\cite{FG}.

Less is known about a connection between nonpolynomial first integrals of the geodesic equation and generalised Killing vectors and tensors. In a series of interesting works \cite{C}--\cite{AHT}, the case of a rational constant of the motion was studied. Demanding the ratio $\frac{\xi^{i_1 \dots i_n}(x) p_{i_1} \dots p_{i_n}}{ \eta^{j_1 \dots j_m}(x) p_{j_1} \dots p_{j_m}}:=\frac{\left(\xi^{(n)},p\right)}{\left(\eta^{(m)},p\right)}$ to be conserved along a geodesic curve
\be\label{ri}
\left(\eta^{(m)},p\right) \{\left(\xi^{(n)},p\right),H \}-\left(\xi^{(n)},p\right) \{\left(\eta^{(m)},p\right),H\}=0,
\ee
one gets the intertwining relation
\be\label{ir}
\eta^{(i_1\dots i_m}\nabla^{i_{m+1}}\xi^{i_{m+2}\dots i_{n+m+1})}-\xi^{(i_1\dots i_n}\nabla^{i_{n+1}}\eta^{i_{n+2}\dots i_{n+m+1})}=0.
\ee
Rewriting (\ref{ri}) in the equivalent form
\be
\frac{\{\left(\xi^{(n)},p\right),H \}}{\left(\xi^{(n)},p\right) }=\frac{\{\left(\eta^{(m)},p\right),H\}}{\left(\eta^{(m)},p\right)}:=(h^{(1)},p),
\ee
where $h^i (x)$ is the so called cofactor of $\xi$ and $\eta$ \cite{AHT},
one can introduce the concept of a Killing pair $(\xi,\eta)$ specified by \cite{C}
\be\label{KP}
\nabla^{(i_1} \xi^{i_2 \dots i_{n+1})}=h^{(i_1} \xi^{i_2 \dots i_{n+1})}, \qquad \nabla^{(i_1} \eta^{i_2 \dots i_{m+1})}=h^{(i_1} \eta^{i_2 \dots i_{m+1})}.
\ee
In particular, the integrability conditions were studied in \cite{C}, while \cite{COD,AHT} provided some explicit examples.
Note that if the cofactor vector field is a gradient of some function, $h_i=-\partial_i \ln{f}(x)$, then the products $\left(\xi^{(n)},p\right)f(x) $ and $\left(\eta^{(m)},p\right)f(x) $ commute with the Hamiltonian and, hence, $\frac{\left(\xi^{(n)},p\right)}{\left(\eta^{(m)},p\right)}=\frac{\left(\xi^{(n)},p\right) f(x)}{\left(\eta^{(m)},p\right) f(x)}$ is functionally dependent on two polynomial first integrals \cite{AHT}.

It is natural to wonder how the situation described above is altered when a first integral of the geodesic equation is represented by a transcendental function on a phase space.
The goal of this work is to construct some metrics admitting nonpolynomial first integrals of the geodesic equation and to reveal possible generalisations of eqs. (\ref{ir}) and (\ref{KP}).

As demonstrated below, the presence of a nonpolynomial first integral of the geodesic equation manifests itself in a chain of vector fields on a curved manifold which obey a specific nonlinear intertwining relation. A Hamiltonian counterpart of the latter, which is obtained by contracting each free index with a canonical momentum, admits an integrating multiplier. The intertwining equation can be resolved by introducing cofactor vector fields, which give rise to generalised Killing equations.

All examples in this paper derive from finite--dimensional real Lie algebras and the whole construction goes in parallel with a group--theoretic description of the Euler top. Given a real Lie algebra with generators $J_i$, $i=1,\dots,n$, the structure relations
\be
[J_i,J_j]=c^k_{ij} J_k, \qquad c^k_{ij}=-c^k_{ji}, \qquad c^p_{ij} c^s_{kp}+c^p_{jk} c^s_{ip}+c^p_{ki} c^s_{jp}=0,
\ee
and an invariant element $\mathcal{I}(J)$
\be\label{In}
[\mathcal{I}(J),J_i ]=0,
\ee
one first introduces the (degenerate) Poisson bracket $\{J_i,J_j\}=c^k_{ij} J_k$ and the quadratic Hamiltonian
\be\label{HAM}
H=\frac 12 \sum_{i=1}^n a_i^2 J^2_i,
\ee
where $a_i$ are real constants (moments of inertia).

As the next step, one considers canonical pairs $(q^i,\pi_i)$, with $i=1,\dots,n$, which obey the Poisson bracket $\{q^i,\pi_j \}={\delta_j}^i$, and constructs a natural phase space realisation of the algebra at hand
\be\label{J}
J_i=c^k_{ij} q^j \pi_k.
\ee
That such $J_i$ obeys the structure relations $\{J_i,J_j\}=c^k_{ij} J_k$ follows from the Jacobi identity $c^p_{ij} c^s_{kp}+c^p_{jk} c^s_{ip}+c^p_{ki} c^s_{jp}=0$. Finally, one substitutes $J_i$ into the Hamiltonian (\ref{HAM}) and regards the latter as the geodesic Hamiltonian for which $\mathcal{I}(J)$ in (\ref{In}) provides a constant of the motion. In general,
not all of the variables $(q^i,\pi_i)$ contribute to $J_i$. Before constructing the geodesic Hamiltonian, one should implement a reduction over cyclic variables. In particular, focusing on $su(2)$ algebra, for which $\{J_i,J_j\}=\epsilon_{ijk}J_k$, where $i=1,2,3$ and $\epsilon_{ijk}$ is the Levi--Civita symbol, one uncovers the Hamiltonian formulation of the Euler top.

The work is organised as follows.

In Sect. 2, we consider three--dimensional real Lie algebras in accord with the Bianchi classification and focus on three instances which are characterised by a transcendental invariant element. Two--dimensional Riemannian metrics are constructed, which admit a generalised Killing pair. As compared to a rational first integral discussed above, the right hand sides of the equations in (\ref{KP}) may involve both $\xi$ and $\eta$, while the intertwining relation (\ref{ir}) is modified accordingly. It is shown that in each case the metric is generated by the cofactor vector field $h^i$ and the geodesic motion is Liouville integrable.

A three--dimensional Riemannian metric is built in Sect. 3 along similar lines. It is demonstrated that a generalised Killing triplet associated with it obeys two intertwining relations, one of which is quadratic in Killing fields and their covariant derivatives, while the other is cubic. Both equations can be resolved by introducing a single cofactor vector field. The corresponding geodesic flow is shown to be Liouville integrable.

In Sect. 4, a specific four--dimensional Riemannian metric is discussed. It admits a generalised Killing triplet obeying a single intertwining equation, resolving of which requires introducing two cofactor vector fields.
In this case, the geodesic Hamiltonian fails to qualify for describing a Liouville integrable system.

In the concluding Sect. 5, we summarise our results and discuss possible further developments. It is argued that Riemannian metrics constructed in this work can be extended to Lorentzian metrics defined on a spacetime involving two extra dimensions. This makes the whole picture more realistic.

\vspace{0.5cm}

\noindent
{\bf 2. Two--dimensional examples}\\

\noindent
Classification of three-dimensional real Lie algebras was accomplished by Bianchi (for a modern exposition see \cite{DNF}). The available options are displayed in Table 1 below, where $\alpha$ designates an arbitrary real constant. In this section, we focus on three instances which are characterised by a transcendental invariant elements $\mathcal{I}(J)$.

Our first example derives from the type-IV algebra. Introducing coordinates $q^a=(z,y,x)$ and momenta $\pi_a=(p_z,p_y,p_x)$ obeying the Poisson bracket $\{q^a,\pi_b \}={\delta_b}^a$, taking into account the structure constants displayed in Table 1, and evaluating $J_a=c^l_{ab} q^b \pi_l$, one gets
\be\label{gen1}
J_1=x p_x+y(p_x+p_y), \qquad J_2=-z(p_x+p_y), \qquad J_3=-z p_x.
\ee
Because $p_z$ does not contribute to (\ref{gen1}), it is a cyclic variable. One can implement a reduction in which $z=-1$, $p_z=0$. Substituting the resulting generators into (\ref{HAM}), one gets a two--dimensional dynamical system governed by the Hamiltonian\footnote{It is straightforward to verify that two out of three constants $(a_1,a_2,a_3)$ in (\ref{HAM}) can be removed by rescaling coordinates and discarding an overall factor.}
\be\label{ham1}
H=\frac 12 g^{ij}p_i p_j=\frac 12 \left(1+\kappa^2+{\left( x+y\right)}^2\right) p_x^2+\frac 12 \left(1+y^2\right) p_y^2+\left(1+ y(x+y)\right) p_x p_y,
\ee
\newpage
\begin{center}
Table 1. The Bianchi classification of three-dimensional real Lie algebras
\end{center}
\begin{eqnarray*}
\footnotesize
\begin{array}{|l|r|r|r|r|r|r|r|r|c|}
\hline
  & \{J_1,J_2 \}  & \{J_1,J_3 \}   & \{J_2,J_3 \}  & \mbox{invariant element}~ \mathcal{I}(J) \\
  \hline
~ \mbox{type I} & 0 & 0 & 0 &  J_1, J_2, J_3 \\
\hline
~ \mbox{type II} & 0 & 0 & J_1 &  J_1 \\
\hline
\mbox{ type III} & J_2-J_3  & -J_2+J_3 & 0 & J_2+J_3\\
\hline
\mbox{ type IV} & J_2+J_3 & J_3 & 0 & \frac{J_2}{J_3}-\ln{J_3} \\
\hline
\mbox{ type V} & J_2 & J_3 & 0 & \frac{J_2}{J_3}\\
\hline
\mbox{ type VI} & \alpha J_2-J_3 & -J_2+\alpha J_3 & 0 & J_3^2 {\left(1+\frac{J_2}{J_3} \right)}^{1+\alpha} {\left(1-\frac{J_2}{J_3} \right)}^{1-\alpha} \\
\hline
~ \mbox{type $$VI$_0$} & 0 & J_2 & J_1& J_1^2-J_2^2  \\
\hline
 \mbox{ type VII} & \alpha J_2+J_3 & -J_2+\alpha J_3 & 0 & (J_2^2 + J_3^2) e^{-2 \alpha \arctan{\frac{J2}{J3}}}  \\
\hline
\mbox{ type $$VII$_0$} & 0 & -J_2& J_1 & J_1^2+J_2^2  \\
\hline
\mbox{ type VIII} & -J_3& -J_2 & J_1 & J_1^2+J_2^2-J_3^2  \\
\hline
 \mbox{ type IX} & J_3 & -J_2 & J_1 & J_1^2+J_2^2+J_3^2  \\
\hline
\end{array}
\end{eqnarray*}
where $\kappa$ is a constant and $p_i=(p_x,p_y)$. The model possesses the integral of motion
\be\label{inv}
\mathcal{I}=p_y/p_x-\ln{p_x},
\ee
which derives from the invariant element in Table 1, and, hence, it is Liouville integrable.

As the next step, one regards (\ref{ham1}) as the geodesic Hamiltonian and the inverse of $g^{ij}$ is used to define the metric tensor
\be
ds^2=g_{ij} dx^i dx^j=\frac{\left(1+y^2\right)dx^2+\left(1+\kappa^2+{(x + y)}^2 \right)dy^2-2\left(1+y (x+y) \right)dx dy}{x^2+\kappa^2 \left(1+y^2\right)},
\ee
where $dx^i=(dx,dy)$.

Introducing two vector fields
\be\label{xieta1}
\xi=\partial_x, \qquad \eta=\partial_y,
\ee
which are prompted by the constituents $J_2$, $J_3$ entering the invariant element $\mathcal{I}$ in Table 1, and analysing the equation $\{ \mathcal{I},H\}=0$, one obtains the intertwining relation
\be\label{intr}
\xi^{(i} \nabla^j \eta^{k)}-\eta^{(i} \nabla^j \xi^{k)}-\xi^{(i} \nabla^j \xi^{k)}=0.
\ee

Then one can try to resolve (\ref{intr}) by turning to the decompositions
\be\label{try}
\nabla^{(i} \xi^{j)}=b_1 h^{(i} \xi^{j)}+b_2 h^{(i} \eta^{j)}, \qquad \nabla^{(i} \eta^{j)}=c_1 h^{(i} \xi^{j)}+c_2 h^{(i} \eta^{j)},
\ee
where $b_1$, $b_2$, $c_1$, $c_2$ are constants and $h^i$ is a cofactor vector field to be fixed below. Substituting (\ref{try}) into (\ref{intr}), one fixes the constants
\be\label{SUPP}
\nabla^{(i} \xi^{j)}=h^{(i} \xi^{j)}, \qquad  \nabla^{(i} \eta^{j)}=h^{(i} \eta^{j)}+h^{(i} \xi^{j)},
\ee
while a direct inspection of (\ref{SUPP}) gives
\be\label{cof1}
h=-(x+y)\partial_x-y \partial_y.
\ee
Note that, if the last term in (\ref{intr}) and the last term entering the rightmost relation in (\ref{SUPP}) were missing, the equations would precisely fit the definition of a Killing pair in \cite{C}. It seems natural to regard $(\xi,\eta)$ obeying eqs. (\ref{intr}), (\ref{SUPP}), (\ref{cof1}) as an example of a generalised Killing pair.

Contracting (\ref{intr}) with $p_i p_j p_k$, one obtains its Hamiltonian counterpart
\be\label{HA}
\left(\xi^{(1)},p \right) \{\left(\eta^{(1)},p \right),H \} -
\left(\eta^{(1)},p \right) \{\left(\xi^{(1)},p \right),H \}-\left(\xi^{(1)},p \right) \{\left(\xi^{(1)},p \right),H \}=0.
\ee
Being multiplied with
\be
\mu=\frac{1}{{\left(\xi^{(1)},p \right)}^2}=\frac{1}{p_x^2},
\ee
eq. (\ref{HA}) can be put into the total derivative form
\be
\frac{d}{d s}\left( p_y/p_x-\ln{p_x} \right)=\{p_y/p_x-\ln{p_x},H\}=0,
\ee
where $s$ is the proper time parameter.

Our second example derives from the Bianchi type--VI algebra. Proceeding as above, one first constructs a realisation in a four--dimensional phase space parametrised by the canonical pairs $(x,p_x)$ and $(y,p_y)$
\be
J_1=\alpha(x p_x+y p_y)-(x p_y+y p_x), \qquad J_2=-p_x+\alpha p_y, \qquad J_3=\alpha p_x-p_y,
\ee
where $\alpha\ne 1$ is a real parameter (see Table 1). These generators give rise to the geodesic Hamiltonian\footnote{For the case at hand, one moment of inertia can be removed by rescaling the coordinates. We assume that $\kappa$ and $\lambda$ do not vanish simultaneously.}
\bea\label{ex2}
&&
H=\frac 12 \left( {(\alpha x-y)}^2 +\kappa^2+\alpha^2 \lambda^2 \right) p_x^2+\frac 12 \left( { (\alpha y-x)}^2+\lambda^2+\alpha^2 \kappa^2\right) p_y^2
\nonumber\\[2pt]
&&
\qquad
+\left((\alpha x-y)(\alpha y-x)-\alpha(\kappa^2+\lambda^2) \right) p_x p_y,
\eea
where $\kappa$ and $\lambda$ are constant parameters,
and the metric tensor
\bea\label{m2}
&&
ds^2=\frac{\left({(\alpha y-x)}^2 +\lambda^2+\alpha^2 \kappa^2 \right) dx^2 +\left({(\alpha x-y)}^2 +\kappa^2+\alpha^2 \lambda^2 \right)dy^2}{{\left(\alpha^2-1 \right)}^2 \left(\lambda^2 y^2 +\kappa^2( \lambda^2+x^2) \right)}
\nonumber\\[2pt]
&&
\qquad
-\frac{2 \left((\alpha x-y)(\alpha y-x)-\alpha\left(\kappa^2+\lambda^2\right) \right)dx dy}{{\left(\alpha^2-1 \right)}^2 \left(\lambda^2 y^2 +\kappa^2( \lambda^2+x^2) \right)}.
\eea

Introducing two vector fields
\be\label{vf}
\xi=-\partial_x+\alpha \partial_y, \qquad \eta=\alpha \partial_x-\partial_y,
\ee
whose form is suggested by $J_2$, $J_3$ above and the invariant element $\mathcal{I}$ in Table 1, one can verify that they satisfy the equation
\be\label{ir2}
-\xi^{(i} \nabla^j \xi^{k)}+\eta^{(i} \nabla^j \eta^{k)}-\alpha\xi^{(i} \nabla^j \eta^{k)}+\alpha\eta^{(i} \nabla^j \xi^{k)}=0.
\ee
Adopting the decompositions similar to (\ref{try}), one can resolve (\ref{ir2})
\be
\nabla^{(i} \xi^{j)}=h^{(i} \eta^{j)}-\alpha h^{(i} \xi^{j)}, \qquad  \nabla^{(i} \eta^{j)}=h^{(i} \xi^{j)}-\alpha h^{(i} \eta^{j)},
\ee
where the cofactor vector field $h$ reads
\be
h=(\alpha x-y) \partial_x+(\alpha y-x) \partial_y.
\ee

Contracting the intertwining relation (\ref{ir2}) with $p_i p_j p_k$ and multiplying the result by the integrating multiplier
\be
\mu=\frac{2}{\alpha^2-1} {\left( \frac{p_x+p_y}{p_x-p_y} \right)}^{\alpha},
\ee
one gets a nonpolynomial constant of the motion
\be
\mathcal{I }={(p_x+p_y)}^{1+\alpha} {(p_x-p_y)}^{1-\alpha}, \qquad \{\mathcal{I},H \}=0,
\ee
which renders the system Liouville integrable.

Our last two--dimensional example relies upon the Bianchi type--VII algebra. In this case the generators read
\be
J_1=\alpha(x p_x+y p_y)-x p_y+y p_x, \qquad J_2=p_x+\alpha p_y, \qquad J_3=\alpha p_x-p_y,
\ee
which give rise to the Hamiltonian
\bea\label{GH3}
&&
H=\frac 12 \left( { \left(\alpha x+y\right)}^2 +\kappa^2+\alpha^2 \lambda^2\right) p_x^2+\frac 12 \left( { \left(\alpha y-x\right)}^2 +\lambda^2+\alpha^2 \kappa^2  \right) p_y^2
\nonumber\\[2pt]
&&
\qquad
+\left((\alpha x+y)(\alpha y-x)+\alpha(\kappa^2-\lambda^2) \right) p_x p_y,
\eea
where $\kappa$ and $\lambda$ are constant parameters.\footnote{Similarly to the Bianchi type--VI case, one of the parameters $(a_1,a_2,a_3)$ entering the Hamiltonian can be removed by rescaling coordinates and discarding an overall factor. We assume that $\kappa$ and $\lambda$ do not vanish simultaneously.}
The corresponding metric differs only slightly from (\ref{m2})
\bea
&&
ds^2=\frac{\left( {(\alpha y-x)}^2+\lambda^2+\alpha^2 \kappa^2\right) dx^2 +\left( {(\alpha x+y)}^2+\kappa^2+\alpha^2 \lambda^2 \right)dy^2}{{\left(\alpha^2+1\right)}^2 \left(\lambda^2 y^2+\kappa^2(\lambda^2+x^2) \right)}
\nonumber\\[2pt]
&&
\qquad
-\frac{2 \left((\alpha x+y)(\alpha y-x)+\alpha(\kappa^2-\lambda^2)\right)dx dy}{{\left(\alpha^2+1\right)}^2 \left( \lambda^2 y^2+\kappa^2(\lambda^2+x^2)\right)}.
\eea

For the case at hand, a generalised Killing pair is formed by
\be
\xi=\partial_x+\alpha \partial_y, \qquad \eta=\alpha \partial_x-\partial_y,
\ee
which obey the intertwining equation
\be\label{INT3}
\xi^{(i} \nabla^j \xi^{k)}+\eta^{(i} \nabla^j \eta^{k)}+\alpha\xi^{(i} \nabla^j \eta^{k)}-\alpha\eta^{(i} \nabla^j \xi^{k)}=0.
\ee
The cofactor vector field $h^i$, which reduces (\ref{INT3}) to
\be
\nabla^{(i} \xi^{j)}=\alpha h^{(i} \xi^{j)}+h^{(i} \eta^{j)}, \qquad  \nabla^{(i} \eta^{j)}=\alpha h^{(i} \eta^{j)}-h^{(i} \xi^{j)},
\ee
has the following form
\be
h=-(\alpha x+y) \partial_x-(\alpha y-x) \partial_y.
\ee

Being contracted with $p_i p_j p_k$, eq. (\ref{INT3}) admits the integrating multiplier
\be
\mu=\frac{2}{\alpha^2+1} e^{-2 \alpha \arctan{\left(\frac{\alpha p_y+p_x}{\alpha p_x-p_y}\right)}},
\ee
which leads to the integral of motion of the geodesic Hamiltonian (\ref{GH3})
\be
\mathcal{I }=e^{-2 \alpha \arctan{\left( \frac{\alpha p_y+p_x}{\alpha p_x-p_y}\right) }} \left(p_x^2+p_y^2 \right).
\ee

It is noteworthy that the presence of a nonpolynomial first integral of the geodesic equation manifests itself in a pair of vector fields on a curved manifold which obey a specific nonlinear intertwining equation. A Hamiltonian counterpart of the latter, which is obtained by contracting each free index with a canonical momentum, admits an integrating multiplier. In each case the intertwining equation can be resolved by introducing a cofactor vector field, which gives rise to generalised Killing equations. It seems natural to consider a set of vector fields entering the intertwining equation as forming a generalised Killing chain.

Concluding this section, we note that for all three examples above the metrics have signature $(+,+)$ and neither the Riemann tensor, nor the Ricci tensor, nor the scalar curvature vanish for generic values of the parameters $\alpha$, $\kappa$, $\lambda$. Curiously enough, in each case the inverse metric is generated by the cofactor vector field
\be
g^{ij}=h^i h^j+g_0^{ij},
\ee
where $g_0^{ij}$ is a constant symmetric matrix.
Computing $(1-g_{11})(1-g_{22})-g_{12}^2 \ne 0$, one concludes that the metrics cannot be regarded as induced on a two--dimensional surface imbedded in a flat three--dimensional space of signature $(+,+,-)$. For each instance both $\xi^i$, $\eta^i$, and $h^i$ obey the equations
\be
\xi^{[i} \nabla^j \xi^{k]}=0, \qquad \eta^{[i} \nabla^j \eta^{k]}=0, \qquad h^{[i} \nabla^j h^{k]}=0.
\ee
Finally, one can verify that in each case the cofactor $h_i$ cannot be represented as a gradient of some function.

\vspace{0.5cm}

\noindent
{\bf 3. A three--dimensional example}\\

\noindent
Our three--dimensional example stems from a four--dimensional real Lie algebra which is specified by the structure relations (see item $A^{\alpha\beta}_{4,6}$ in \cite{PSWZ})
\be\label{3dal}
[J_1,J_4]=\alpha J_1, \qquad [J_2,J_4]=\beta J_2-J_3, \qquad [J_3,J_4]=J_2+\beta J_3,
\ee
where $\alpha\ne 0$, $\beta \geq 0$ are real parameters. The algebra admits two invariant elements \cite{PSWZ}
\be\label{CE3}
\mathcal{I}_1=\frac{J_1^{\frac{2\beta}{\alpha}} }{J_2^2+J_3^2}, \qquad \mathcal{I}_2=\left(J_2^2+J_3^2 \right) e^{2 \beta \arctan{\frac{J_2}{J_3}}},
\ee
which commute with each generator.

Introducing coordinates $q^a=(z,y,x,w)$ and momenta $\pi_a=(p_z,p_y,p_x,p_w)$ obeying the Poisson bracket $\{q^a,\pi_b\}={\delta_b}^a$ and computing the phase space functions (\ref{J}), one reveals that $p_w$ does not contribute. Setting $w=1$, $p_w=0$, one gets a realisation in a sex--dimensional phase space
\be
J_1=\alpha p_z, \quad J_2=\beta p_y-p_x, \quad J_3=\beta p_x+p_y, \quad J_4=y p_x-x p_y-\beta(x p_x+y p_y)-\alpha z p_z.
\ee

In accord with (\ref{HAM}), the latter can be used to construct a three--dimensional dynamical system which is described by the geodesic Hamiltonian
\bea\label{H4}
&&
H=\frac 12 g^{ij} p_i p_j=\frac 12 \left({(\beta x-y)}^2+\lambda^2+\beta^2 \sigma^2 \right) p_x^2+\frac 12 \left({(\beta y+x)}^2+\sigma^2+\beta^2 \lambda^2 \right) p_y^2
\nonumber\\[2pt]
&&
\qquad \qquad \qquad \quad +\frac 12 \alpha^2 \left(z^2+\kappa^2 \right) p_z^2+\left((\beta x-y)(\beta y+x)-\beta\left(\lambda^2-\sigma^2 \right) \right) p_x p_y
\nonumber\\[2pt]
&&
\qquad \qquad \qquad \quad
+\alpha z (\beta x-y) p_x p_z+\alpha z (\beta y+x)p_y p_z,
\eea
where $p_i=(p_x,p_y,p_z)$ and $\lambda$, $\sigma$, $\kappa$ are constant parameters originating from moments of inertia\footnote{As above, we eliminate one moment of inertia by discarding an overall number coefficient in the Hamiltonian. We assume that $\kappa$, $\lambda$, $\sigma$ do not vanish simultaneously.} in (\ref{HAM}). Two integrals of motion, which follow from the invariant elements (\ref{CE3}), read
\be\label{COM3}
\mathcal{I }_1=\frac{{p_z}^{\frac{2\beta}{\alpha}}}{p_x^2+p_y^2}, \qquad \mathcal{I}_2=\left(p_x^2+p_y^2 \right) e^{2 \beta \arctan{\frac{\beta p_y-p_x}{\beta p_x+p_y}}}.
\ee
Because $(H,\mathcal{I}_1,\mathcal{I}_2)$ are mutually commuting and functionally independent, the system is Liouville integrable.

Switching to the language of the Riemannian geometry, one uses the inverse of $g^{ij}$ to construct the line element
\bea\label{3dmetr}
&&
ds^2=g_{ij}dx^i dx^j=
\frac{\left(\kappa^2 {(\beta y+x)}^2+(\sigma^2+\beta^2 \lambda^2) (z^2+\kappa^2) \right) dx^2} { \Omega(x,y,z)}
\nonumber\\[2pt]
&&
\qquad \quad
+\frac{\left(\kappa^2 {(\beta x-y)}^2+(\lambda^2+\beta^2 \sigma^2) (z^2+\kappa^2) \right) dy^2}{\Omega(x,y,z)}
\nonumber\\[2pt]
&&
\qquad \quad
+\frac{\left(\sigma^2 y^2+\lambda^2 (x^2+\sigma^2) \right) \alpha^{-2} {\left(1+\beta^2 \right)}^2  dz^2}{ \Omega(x,y,z) }
\nonumber\\[2pt]
&&
\qquad \quad
+\frac{2 \left( \kappa^2 (1-\beta^2) x y+\beta \kappa^2 (y^2-x^2)+\beta (\lambda^2-\sigma^2)(z^2+\kappa^2) \right) dx dy}{\Omega(x,y,z)}
\nonumber\\[2pt]
&&
\qquad \quad
+\frac{2 \alpha^{-1} (1+\beta^2) (\sigma^2 y-\beta \lambda^2 x) z dx dz}{\Omega(x,y,z)}-\frac{2 \alpha^{-1} (1+\beta^2) (\lambda^2 x+\beta \sigma^2 y)  z dy dz}{\Omega(x,y,z)},
\eea
where $\Omega(x,y,z)={\left(1+\beta^2 \right)}^2 \left(\kappa^2(\lambda^2 x^2+\sigma^2 y^2)+\sigma^2 \lambda^2(z^2+\kappa^2) \right)$.

Introducing three vector fields
\be
\xi=\partial_x, \qquad \eta=\partial_y, \qquad \mu=\partial_z,
\ee
which are suggested by the constituents $p_x=\left(\xi^{(1)},p \right)$,  $p_y=\left(\eta^{(1)},p \right)$,  and $p_z=\left(\mu^{(1)},p \right)$ entering constants of the motion (\ref{COM3}), and analysing $\{\mathcal{I}_{1,2},H \}=0$, one obtains two intertwining relations
\bea\label{ENT3}
&&
\xi_{(i} \nabla_j \xi_{k)}+\eta_{(i} \nabla_j \eta_{k)}+\beta \xi_{(i} \nabla_j \eta_{k)}-\beta \eta_{(i} \nabla_j \xi_{k)}=0,
\nonumber\\[4pt]
&&
\beta \xi_{(i}\xi_j \nabla_k \mu_{l)}+\beta \eta_{(i}\eta_j \nabla_k \mu_{l)}-\alpha \xi_{(i}\mu_j \nabla_k \xi_{l)}-\alpha \eta_{(i}\mu_j \nabla_k \eta_{l)}=0,
\eea
where $\alpha$, $\beta$ are parameters entering the algebra (\ref{3dal}). Note that the second equation in (\ref{ENT3}) is cubic in the fields and their covariant derivatives.

Similarly to the examples in the preceding section, it seems natural to try the decompositions
\bea\label{dec3}
&&
\nabla^{(i} \xi^{j)}=b_1 h^{(i} \xi^{j)}+b_2 h^{(i} \eta^{j)}+b_3 h^{(i} \mu^{j)},
\nonumber\\[2pt]
&&
\nabla^{(i} \eta^{j)}=c_1 h^{(i} \xi^{j)}+c_2 h^{(i} \eta^{j)}+c_3 h^{(i} \mu^{j)},
\nonumber\\[2pt]
&&
\nabla^{(i} \mu^{j)}=d_1 h^{(i} \xi^{j)}+d_2 h^{(i} \eta^{j)}+d_3 h^{(i} \mu^{j)},
\eea
where $b_i$, $c_i$, $d_i$, with $i=1,2,3$, are constants to be fixed from (\ref{ENT3}) and $h^i$ is the cofactor vector field to be determined from (\ref{dec3}). A straightforward computation yields
\be
\nabla^{(i} \xi^{j)}=\beta h^{(i} \xi^{j)}+h^{(i} \eta^{j)}, \qquad  \nabla^{(i} \eta^{j)}=\beta h^{(i} \eta^{j)}-h^{(i} \xi^{j)}, \qquad \nabla^{(i} \mu^{j)}=\alpha h^{(i} \mu^{j)},
\ee
where
\be
h=-(\beta x-y)\partial_x-(\beta y+x) \partial_y-\alpha z \partial_z,
\ee
Computing $\partial_i h_j-\partial_j h_i$, one can verify that
$h_i$ is not a gradient of a scalar function. By construction, the Hamiltonian counterpart of each intertwining relation in (\ref{ENT3}) admits an integrating multiplier and, hence, $(\xi,\eta,\mu)$ form a generalised Killing triplet.

Concluding this section, we note that the metric (\ref{3dmetr}) has signature $(+,+,+)$ and neither the Riemann tensor, nor the Ricci tensor, nor the scalar curvature vanish for generic values of the parameters $\alpha$, $\beta$, $\lambda$, $\sigma$, $\kappa$. Interestingly enough, similarly to the two--dimensional examples, the inverse metric in (\ref{H4}) is generated by the cofactor vector field $g^{ij}=h^i h^j+g_0^{ij}$, where $g_0^{ij}$ is a symmetric constant matrix.

\vspace{0.5cm}
\noindent
{\bf 4. A four--dimensional example}\\

\noindent
A peculiar feature of the example in the preceding section is that the generalised Killing triplet $(\xi,\eta,\mu)$ obeys two intertwining relations, which can be resolved by introducing a single cofactor vector field $h$. Our goal in this section is to discuss the opposite situation in which resolving a single intertwining equation requires introducing two cofactor vector fields.

Let us consider the five--dimensional real Lie algebra (see item $A_{5,30}$ in \cite{PSWZ})
\begin{align}\label{al4}
&
[J_2,J_4]=J_1, && [J_3,J_4]=J_2, && [J_1,J_5]=(\alpha+1) J_1,
\nonumber\\[2pt]
&
[J_2,J_5]=\alpha J_2, && [J_3,J_5]=(\alpha-1)J_3, && [J_4,J_5]=J_4,
\end{align}
where $\alpha\ne -1,0,1$ is a real parameter, which admits a single invariant element \cite{PSWZ}
\be\label{inv5}
\mathcal{I}=J_1^{-2\alpha} {\left(J_2^2-2 J_1 J_3 \right)}^{\alpha+1}.
\ee

Proceeding as above, one first builds a realisation of (\ref{al4}) in an eight--dimensional phase space
\bea\label{real4}
&&
J_1=(1+\alpha)p_w, \qquad \qquad J_2 =\alpha p_z+x p_w, \qquad \qquad J_3=x p_z+(\alpha-1) p_y,
\nonumber\\[2pt]
&&
J_4=p_x-y p_z-z p_w, \qquad J_5=y p_y-x p_x-\alpha (y p_y+z p_z)-(\alpha+1) w p_w,
\eea
where $(x,p_x)$, $(y,p_y)$, $(z,p_z)$, and $(w,p_w)$ are canonically conjugate pairs obeying the conventional Poisson brackets.\footnote{When obtaining eq. (\ref{real4}), one first introduces coordinates $q^i=(w,z,y,x,f)$ and momenta $\pi_i=(p_w,p_z,p_y,p_x,p_f)$, then computes (\ref{J}) and reveals that $p_f$ does not contribute. Implementing the reduction $f=1$, $p_f=0$, one arrives at (\ref{real4}).} Then one constructs the Hamiltonian (see eq. (\ref{HAM}))
\bea\label{Ham5}
&&
H=\frac 12 \left(x^2+\lambda^2\right) p_x^2+\frac 12 {(\alpha-1)}^2 \left(y^2+\sigma^2\right) p_y^2+\frac 12 \left(\sigma^2 x^2 + \lambda^2 y^2 + \alpha^2 (\kappa^2 + z^2) \right) p_z^2
\nonumber\\[2pt]
&&
\qquad +\frac 12 \left(\kappa^2 x^2 + \lambda^2 z^2+ {(\alpha+1)}^2 (w^2 +\rho^2) \right) p_w^2+(\alpha-1) x y p_x p_y+(\alpha x z -\lambda^2 y) p_x p_z
\nonumber\\[2pt]
&&
\qquad +\left((\alpha+1)x w  - \lambda^2 z \right) p_x p_w+ (\alpha-1) \left(\sigma^2 x + \alpha y z \right)p_y p_z+(\alpha^2-1) y w p_y p_w
\nonumber\\[2pt]
&&
\qquad
+(\alpha \kappa^2 x + \alpha(1+\alpha)z w  + \lambda^2 y z) p_z p_w,
\eea
where $(\lambda,\sigma,\kappa,\rho)$ are constants, for which the invariant (\ref{inv5}) provides the integral of motion
\be\label{inv6}
\mathcal{I}={\left({(\alpha p_z +x p_w)}^2-2(\alpha+1) p_w (x p_z+(\alpha-1) p_y)\right)}^{\alpha+1} p_w^{-2 \alpha}.
\ee
Note that the resulting model does not qualify for a Liouville integrable system as it involves four degrees of freedom and only two integrals of motion. In what follows, we assume that $\kappa$, $\lambda$, $\sigma$, $\rho$ do not vanish simultaneously.

Treating (\ref{Ham5}) as the geodesic Hamiltonian $H=\frac 12 g^{ij} p_i p_j$, with $p_i=(p_x,p_y,p_z,p_w)$, one can build the metric $ds^2=g_{ij} dx^i dx^j$. Unfortunately, its explicit form is too unwieldy to be presented here. Yet, a generalised Killing triplet, which derives from (\ref{inv6}), is quite readable. So, in what follows, we concentrate on it.

The constituents $J_1$, $J_2$, $J_3$ entering (\ref{inv5}), as well as their realisation by means of eqs. (\ref{real4}) and (\ref{inv6}), suggest introducing three vector fields
\be\label{ktr}
\xi=(\alpha+1)\partial_w, \qquad \eta=\alpha \partial_z+x \partial_w, \qquad \mu=(\alpha-1)\partial_y+x \partial_z,
\ee
while $\{\mathcal{I},H \}=0$ yields the intertwining equation
\be\label{ir4}
(\alpha+1)\xi^{(i} \eta^j \nabla^k \eta^{l)}- (\alpha+1)\xi^{(i} \xi^j \nabla^k \mu^{l)}-\alpha \eta^{(i} \eta^j \nabla^k \xi^{l)}+(\alpha-1)\xi^{(i} \mu^j \nabla^k \xi^{l)}=0,
\ee
which is cubic in the fields and their covariant derivatives.

At this stage, one could try to resolve (\ref{ir4}) by using decompositions similar to (\ref{dec3}).
In our examples above, it was always possible to express all constants entering the decompositions in terms of a single member of the set, while the latter could be removed by rescaling $h^i$.
For the case at hand, a straightforward computation shows that two of nine constants entering (\ref{dec3}) survive, which means that the triplet $(\xi^i,\eta^i,\mu^i)$ actually requires introducing two cofactor vector fields. Indeed,
considering
\bea\label{cof}
&&
h_1=-(1+\alpha)x \partial_x+(1-\alpha^2) y \partial_y-\alpha(1+\alpha) z \partial_z-{(1+\alpha)}^2 w \partial_w,
\nonumber\\[2pt]
&&
h_2=\lambda^2 \left(\partial_x-y \partial_z-z \partial_w \right),
\eea
one can establish the relations
\bea\label{gke4}
\nabla^{(i} \xi^{j)}=h_1^{(i} \xi^{j)}, \qquad \nabla^{(i} \eta^{j)}=\frac{\alpha}{\alpha+1} h_1^{(i} \eta^{j)}+h_2^{(i} \xi^{j)}, \qquad
\nabla^{(i} \mu^{j)}=\frac{\alpha-1}{\alpha+1} h_1^{(i} \mu^{j)}+h_2^{(i} \eta^{j)},
\eea
which, in their turn, resolve (\ref{ir4}). By construction, the Hamiltonian counterpart of eq. (\ref{ir4})
admits an integrating multiplier and, hence, (\ref{ktr}) describes a generalised Killing triplet. Interestingly enough, the generalised Killing equations (\ref{gke4}) involve two cofactor vector fields.

Note that neither $h_1$ nor $h_2$ can be represented as a gradient of a scalar function.
In contrast to the examples in the preceding sections, it appears problematic to link $g^{ij}$ to $h_1^i$ and $h_2^i$.

Further examples related to real Lie algebras can be constructed along similar lines by making use of the results in \cite{PSWZ}.

\vspace{0.5cm}
\noindent
{\bf 5. Conclusion}\\

\noindent
Summarising our consideration above, it seems reasonable to regard a set of tensor fields as forming a generalised Killing chain, if they satisfy an intertwining equation, involving the fields and their covariant derivatives, such that its Hamiltonian counterpart admits an integrating multiplier.
The intertwining equation is assumed to be symmetric in external indices and the Hamiltonian counterpart is obtained by contracting each index with a canonical momentum $p_i$. The Hamiltonian counterpart admits an integrating multiplier if, being multiplied with a specific scalar function and restricted to a geodesic curve, it can be cast into a total derivative form. In particular, the rational case discussed in the Introduction (see eq. (\ref{ri})) admits the integrating multiplier ${\left(\xi^{(n)},p\right)}^{-1} {\left(\eta^{(m)},p\right) }^{-1}$.

All metrics constructed above are Riemannian. Introducing two extra dimensions para\-metrised by the double null coordinates $t$ and $v$ and modifying the metric
\be
ds^2 \quad \to \quad ds^2-2 dt dv,
\ee
one obtains a Lorentzian counterpart which preserves symmetries of the original metric and possesses an extra covariantly constant null Killing vector field
$\partial_v$ (for more details and further references see \cite{GHKW}).

As a possible continuation of this work, it would be interesting to formulate necessary and sufficient conditions for the existence of a generalised Killing chain without invoking the Hamiltonian formalism.

As was demonstrated above, in some cases a metric admitting a nonpolynomial first integral of the geodesic equation is generated by a cofactor vector field entering generalised Killing equations. A geometric meaning of such an interrelationship is worth studying further.

\vspace{0.5cm}

\noindent{\bf Acknowledgements}\\

\noindent
This work is supported by the Russian Foundation for Basic Research, grant No 20-52-12003.


\begin{thebibliography}{nn}
\bibitem{FKK}
V. Frolov, P. Krtous, D. Kubiznak, {\it Black holes, hidden symmetries, and complete integrability}, Living Rev. Rel. {\bf 20} (2017) 6, arXiv:1705.05482.
\bibitem{GHKW}
G.W. Gibbons, T. Houri, D. Kubiznak, C. Warnick, {\it Some spacetimes with higher rank Killing--Stackel tensors}, Phys. Lett. B {\bf 700} (2011) 68, arXiv:1103.5366.
\bibitem{GR}
G.W. Gibbons, C. Rugina, {\it  	
Goryachev--Chaplygin, Kovalevskaya, and Brdi\v{c}ka--Eardley--Nappi--Witten pp--waves spacetimes with higher rank St\"ackel--Killing tensors }, J. Math. Phys. {\bf 52} (2011) 122901, arXiv:1107.5987.
\bibitem{G}
A. Galajinsky, {\it Higher rank Killing tensors and Calogero model}, Phys. Rev. D {\bf 85} (2012) 085002, arXiv:1201.3085.
\bibitem{MCG}
M. Cariglia, G.W. Gibbons, {\it Generalised Eisenhart lift of the Toda chain}, J. Math. Phys. {\bf 55} (2014) 022701, arXiv:1312.2019.
\bibitem{CGHHKZ}
M. Cariglia, G.W. Gibbons, J.W. van Holten, P.A. Horv\'athy, P. Kosinski, P.M. Zhang, {\it Killing tensors and canonical geometry}, Class. Quant. Grav. {\bf 31} (2014) 125001, arXiv:1401.8195.
\bibitem{CG}
M. Cariglia, A. Galajinsky, {\it Ricci-flat spacetimes admitting higher rank Killing tensors}, Phys. Lett. B {\bf 744} (2015) 320, arXiv:1503.02162.
\bibitem{GF}
S. Filyukov, A. Galajinsky, {\it Self-dual metrics with maximally superintegrable geodesic flows}. Phys. Rev. D {\bf 91} (2015) 104020, arXiv:1504.03826.
\bibitem{GM}
A. Galajinsky, I. Masterov, {\it Eisenhart lift for higher derivative systems}, Phys. Lett. B {\bf 765} (2017) 86, arXiv:1611.04294.
\bibitem{G1}
A. Galajinsky, {\it Geometry of the isotropic oscillator driven by the conformal mode}, Eur. Phys. J. C {\bf 78} (2018) 72, arXiv:1712.00742.
\bibitem{CGGH}
M. Cariglia, A. Galajinsky, G.W. Gibbons, P.A. Horv\'athy, {\it Cosmological aspects of the Eisenhart--Duval lift}, Eur. Phys. J. C {\bf 78} (2018) 314, arXiv:1802.03370.
\bibitem{FG} 	
A.P. Fordy, A. Galajinsky, {\it Eisenhart lift of 2--dimensional mechanics}, Eur. Phys. J. C {\bf 79} (2019) 301, arXiv:1901.03699.
\bibitem{C}
C.D. Collinson, {\it A note on the integrability conditions for the existence of rational first integrals of the geodesic equations in a Riemannian space}, Gen. Rel. Grav. {\bf 18} (1986) 207.
\bibitem{COD}
C.D. Collinson, P.J. O'Donnell, {\it A class of empty spacetimes admitting a rational first integral of the geodesic equation}, Gen. Rel. Grav. {\bf 24} (1992) 451.
\bibitem{AHT}
A. Aoki, T. Houri, K. Tomoda, {\it Rational first integrals of geodesic equations and generalised hidden symmetries}, Class. Quant. Grav. {\bf 33} (2016) 195003, arXiv:1605.08955.
\bibitem{DNF}
B.A. Dubrovin, A.T. Fomenko, S.P. Novikov, {\it Modern geometry – methods and applications. Part I. The geometry of surfaces, transformation groups, and fields}. Graduate Texts in Mathematics, Vol. 93, Springer-Verlag, New York, 1984.
\bibitem{PSWZ}
J. Patera, R.T. Sharp, P. Winternitz, H. Zassenhaus, {\it Invariants of real low dimension Lie algebras}, J. Math. Phys. {\bf 17} (1976) 986.
\end{thebibliography}
\end{document}